# The First Hardware Demonstration of a Universal Programmable RRAM-based Probabilistic Computer for Molecular Docking

Yihan He[1†], Ming-Chun Hong[2,3†], Qiming Ding[4], Chih-Sheng Lin[2,3], Chih-Ming Lai[3], Chao Fang[1], Xiao Gong[1], Tuo-Hung Hou[2,5*], and Gengchiau Liang[1,5*]

[1]Department of Electrical and Computer Engineering, National University of Singapore, 117583 Singapore
[2]Department of Electrical Engineering and Institute of Electronics, National Yang-Ming Chiao Tung University, Hsinchu, Taiwan
[3]Electronic and Optoelectronic System Research Laboratories, Industrial Technology Research Institute, Hsinchu, Taiwan
[4]Center on Frontiers of Computing Studies, Peking University, Beijing 100871, China
[5]Industry Academia Innovation School, National Yang-Ming Chiao Tung University, Hsinchu, Taiwan

[†]These authors contribute equally;   [*]Email: thhou@nycu.edu.tw; gcliang@nycu.edu.tw

*Molecular docking is a critical computational strategy in drug design and discovery, but the complex diversity of biomolecular structures and flexible binding conformations create an enormous search space that challenges conventional computing methods. Although quantum computing holds promise for these challenges, it remains constrained by scalability, hardware limitations, and precision issues. Here, we report a prototype of a probabilistic computer (p-computer) that efficiently and accurately solves complex molecular docking for the first time, overcoming previously encountered challenges. At the core of the system is a p-computing chip based upon our artificial tunable probabilistic bits (p-bits), which are compatible with computing-in-memory schemes, based upon 180 nm CMOS technology and BEOL $HfO_2$ RRAM. We successfully demonstrated the superior performance of the p-computer in practical ligand-protein docking scenarios. A 42-node molecular docking problem of lipoprotein with LolA-LolCDE complex—a key point in developing antibiotics against Gram-negative bacteria, was successfully solved. Our results align well with the Protein-Ligand Interaction Profiler tool. This work marks the first application of p-computing in molecular docking-based computational biology, which has great potential to overcome the limitations in success rate and efficiency of current technologies in addressing complex bioinformatics problems.*

## Introduction

In recent years, probabilistic computing (p-computing), as an emerging unconventional computational paradigm, has garnered significant attention due to its unique advantages in solving combinatorial optimization problems (COPs), such as integer factorization[1–5], the traveling salesman problem (TSP)[6,7], and Boolean satisfiability (SAT)[8–10]. The fundamental computing unit of p-computing is the probabilistic bit (p-bit)[4,11–17], which exhibits a tunable sigmoidal stochastic input-output characteristic. A probabilistic computer (p-computer)[18–21] based on this is typically constructed from energy models such as Boltzmann machines and the Ising model. The core advantages of p-computing lie in its ability to efficiently explore solution spaces through approximate Gibbs sampling and to naturally quantify the relative quality of candidate solutions using probability distributions. Furthermore, the tunable stochasticity of p-bits provides an innovative hardware solution for simulating quantum annealing processes under room temperature conditions.

In terms of hardware implementation of p-bits and p-computers, existing CMOS-only solutions predominantly rely on pseudo-random number generators (PRNGs), which incur considerable area and energy overheads[22–24]. To achieve true randomness and high energy efficiency, a concept of hybrid architecture that combining standard CMOS technology with emerging nanotechnologies (CMOS + X) has been proposed[20]. However, current CMOS+X architecture-based p-computers typically employ an ex-situ p-bit[1,6,10,20,25] configuration when addressing hard computational problems. In this setup, stochastic nanodevices (X), are responsible for generating stochastic bitstreams, while external deterministic CMOS circuits handle the subsequent processes of sampling, state storage, state computation and state updates. This disjointed architecture inherently reintroduces the von Neumann memory bottleneck, leading to increased latency, energy inefficiencies and therefore constrains the efficiency and scalability of p-computing systems. Besides, the impact of inherent device-to-device variations, the interactions between stochastic nanodevices and deterministic CMOS circuits, and the potential applications of these hybrid architectures remain areas requiring further investigation.

The revolutionary breakthrough of AlphaFold[26] in protein structure prediction highlights the critical role of computational methods in addressing complex biological problems. Similarly, p-computing, with its unique advantages, is expected to hold promise for tackling challenging problems in computational biology. Among the core problems in computational biology, molecular docking has broad applications in drug design[27,28], enzyme catalytic reaction mechanism research[29], and antibody-antigen recognition[30]. Its goal is to predict the preferred binding conformation of a ligand molecule with a target biomolecule to form a stable complex. However, the structural complexity of biomolecules and the high flexibility of binding modes present significant challenges to this task. Traditional approaches, such as those based on scoring functions[31,32], often rely on approximate models and empirical data, which limit their ability to accurately predict binding conformations. In recent studies, quantum computing methods such as Gaussian Boson Sampling (GBS)[33–36] and Quantum Approximate Optimization Algorithm (QAOA)[37] have provided novel approaches to solving molecular docking problems by reformulating them as clique problems[38–41]. While these methods have shown promise in solving certain docking instances, quantum computing approaches still encounter significant bottlenecks in practical applications. For instance, sparse adjacency matrices, common in molecular docking problems, require carefully fine-tuned experimental configurations to approximate zero parameters, a process that remains highly challenging in experimental implementations. Furthermore, the output of GBS does not guarantee the formation of cliques and additional post-processing algorithms[42–44] are required to optimize solution quality[38,39]. On the other hand, QAOA tends to getting trapped in local energy minima during its evolution[41], making it difficult to converge to the ground state. These limitations substantially restrict the success probability and scalability of quantum computing methods in solving clique problems-based molecular docking.

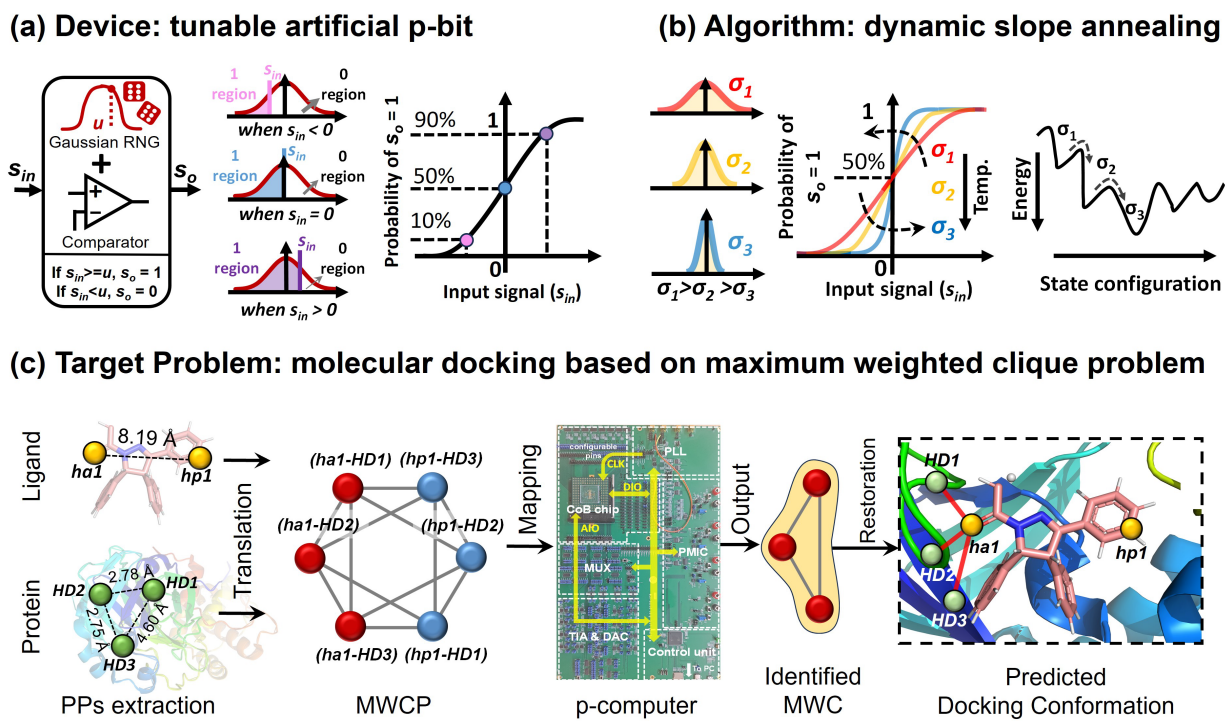

**Fig. 1** Overview of p-computing in computational biology, highlighting the device, algorithm and the graph problem-based molecular docking problem. (a) Concept diagram of the proposed artificial tunable p-bit comprises of a Gaussian RNG and a comparator. The output state ($s_o$) of the p-bit is determined by comparing the accumulated input signal ($s_{in}$) received from interconnected p-bits with a stochastic threshold $u$ generated by the Gaussian RNG. (b) The DSA algorithm dynamically adjusts the standard deviation of the Gaussian RNG, allowing modulation of the stochasticity and altering the steepness of the sigmoidal probability curves. During iterative computation, $\sigma$ is globally tuned to guide the system toward convergence and stability. (c) Schematic representation of solving the protein-ligand docking problem through a MWCP framework, in which the workflow includes pharmacophore points extraction, graph translation, and docking conformation restoration processes. Specifically, the pharmacophore point binding pairs between the ligand and protein will be extracted first and then translated into graph vertices and edges to represent the chemical properties matching and geometric compatibility. Subsequently, the optimal solution of the MWCP solved by p-computer is used to predict binding conformations, which are then restored to docking results for a ligand-protein complex. A more detailed and complete schematic diagram of solving molecular docking under the p-computing framework is provided in Supplementary Figure S1. The example of SARS-CoV-2 Mpro (COVID-19) docking with covalent pyrazoline-based inhibitors PM-2-020B (PDB ID: 8SKH) is used for demonstration.

In this work, we first propose an in-situ artificial p-bit compatible with compute-in-memory (CIM) schemes, combining a resistive random-access memory (RRAM) cell with Gaussian random number generator (RNG) and a comparator (**Fig. 1a**). Unlike conventional stochastic devices in CMOS + X architectures (such as stochastic magnetic tunnel junctions), where the sigmoidal input-output relationship arises from the device's inherent random physical properties and fixed upon fabrication. In contrast, our proposed p-bit design eliminates the dependence on devices' intrinsic stochastic characteristics, effectively lowering the technical barriers for constructing p-bits and providing greater flexibility and adaptability, thus promoting the commercialization of p-computing systems. Leveraging the tunable stochasticity of the proposed artificial p-bit, we further develop a novel dynamic slope annealing (DSA) mechanism (**Fig. 1b**). Unlike conventional simulated annealing (SA), which controls the annealing dynamics by adjusting the global scale for the strength of the interactions like an "inverse temperature", DSA globally modulates the steepness of the p-bit sigmoidal curve by dynamically tuning the standard deviation ($\sigma$) of the Gaussian distribution to locate the global optimum. In practical applications, we developed a computational framework tailored for p-computing to translate the molecular docking into a Maximum Weighted Clique Problem (MWCP) and further reformulated as a quadratic unconstrained binary optimization (QUBO) problem (**Fig. 1c**). Based on this, we fabricated a p-computer prototype for molecular docking by integrating back-end-of-line (BEOL) $HfO_2$ RRAM chips with 180 nm CMOS technology and validated its performance in successfully solving a real 42-node lipoprotein and the LolCDE-LolA complex docking scenario without any postprocessing. To the best of our knowledge, this study represents both the first successful application of p-computing in computational biology and the largest-scale solution of a molecular docking problem based on graph theory.

## Results

### A. P-computing system compatible with CIM and its core artificial tunable p-bit

Our p-computing platform is built upon the architecture of a discrete Hopfield Neural Network (DHNN)[7,45], as shown in **Fig. 2(a)**. DHNN is a single-layer, recurrent neural network characterized by self-feedback and interconnections. This architecture performs optimization by minimizing the following energy function:

$$E = -\frac{1}{2}\sum_{i,j} J_{ij} x_i x_j + \sum_i h_i x_i \qquad (1)$$

where $J_{ij}$ represents the interconnection matrix between p-bits, $h_i$ denotes the local bias of the $i$-th p-bit, and $x_i$ is the instantaneous state of the p-bit, which takes binary values 0 or 1.

However, a standard DHNN encounters challenges when solving complex combinatorial optimization problems, such as the molecular docking problem here, as it is prone to getting trapped in local energy minima, thereby limiting the exploration of global optimal solutions. To address this issue, we introduced an in-RRAM bias with Gaussian RNG into the system, which adjusts the stochasticity of p-bits during the state update process. This enhanced randomness increases the likelihood of escaping local minima, thereby improving the system's capability to explore the solution space globally. During the iterative process for solving the molecular docking problem, the p-bits' states are updated according to the following rules:

$$\sum_j J_{ij} x_j + h_i = s_{in}, \quad \begin{cases} s_{in} \geq u, x_i^{t+1} = 1 \\ s_{in} < u, x_i^{t+1} = 0 \end{cases} \qquad (2)$$

where $s_{in}$ represents the net input to the $i$-th p-bit, and $u$ is the dynamic threshold value provided by the Gaussian RNG, which is taken as an integer.

To implement this p-computing framework in hardware, we adopted $HfO_2$-based RRAM as the core device for this DHNN architecture. RRAM offers high density, low power consumption, and non-volatility, making it an ideal choice for hardware-level p-bit networks.

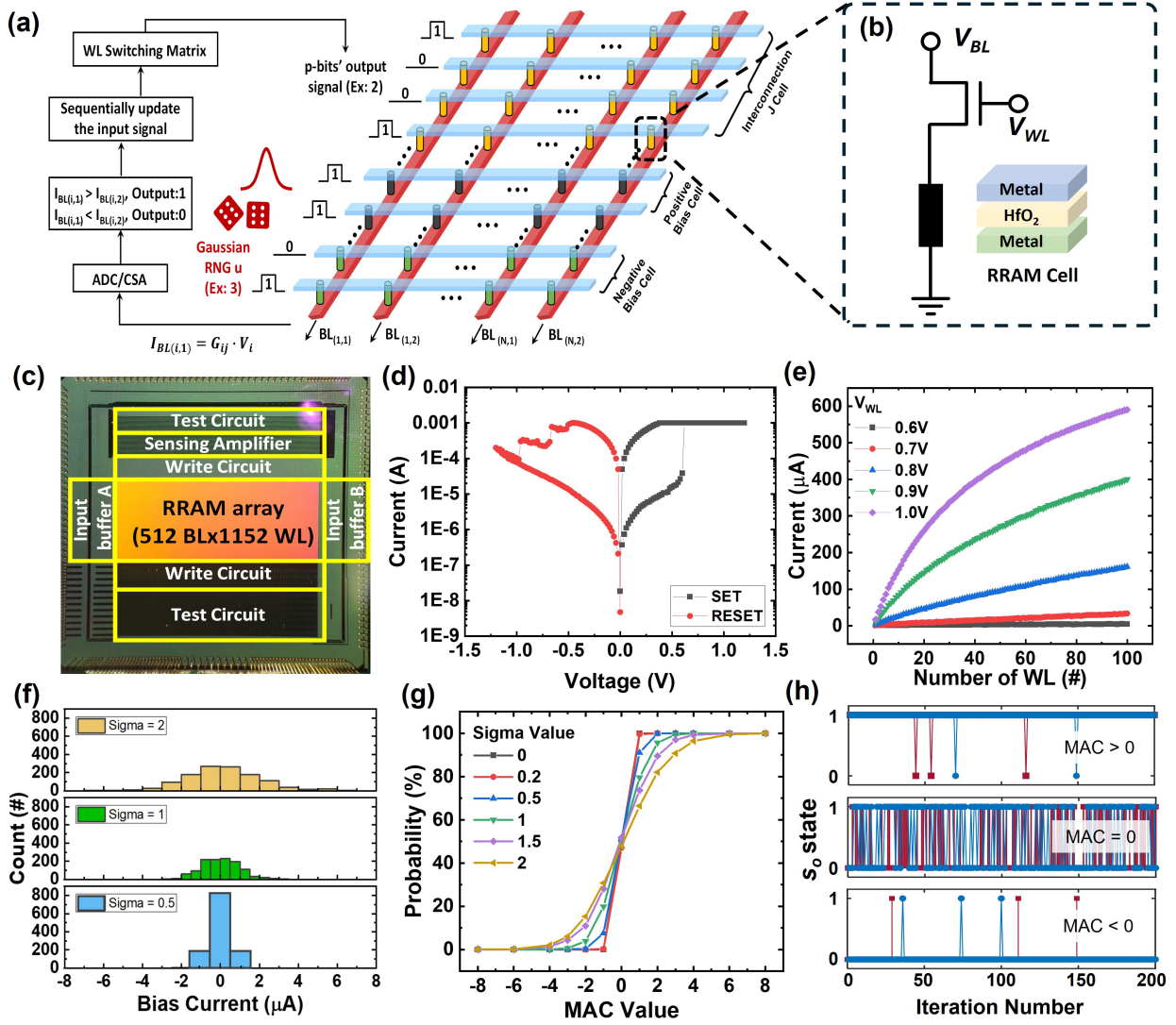

**Fig. 2** Experimental demonstration of the in-memory p-computing system and the artificial p-bit. (a) Schematic of the CIM hardware mapping under the DHNN architecture, in which interconnection values are encoded using pairs of RRAM cells. A p-bit with tunable stochasticity is activated by different numbers of cells in the bias region. (b) 1T1R structure of the DHNN hardware, which needs to be used in conjunction with subsequent Gaussian RNG and comparator circuits. (c) Photograph of the p-computing chip integrating the 1T1R RRAM array with the 180-nm CMOS technology. (d) Characteristics of switching curves of the standalone RRAM cell. (e) Experimentally measured cumulated BL current (MAC output) by activating different numbers of WLs at different WL voltages. (f) Bias current distributions controlled by the sigma value of the Gaussian RNG at $\sigma$ = 0.5, 1, and 2. (g) A series of sigmoidal probability curves of the artificial tunable p-bit (probability of output being 1) under different sigma values ranging from 0 to 2. (h) Time snapshots of the output states for three different inputs (MAC = +4, 0, and −4 respectively) observed experimentally (solid blue line) and by simulation (solid red line) at $\sigma$ = 2.

In this system, the RRAM chip not only serves as a storage element but is also integrated closely with peripheral sensing circuit to generate tunable stochastic output signals, enabling the core functionality of p-bits. **Fig. 2(b)** illustrates the 1T1R array structure of the p-computing chip integrating the RRAM array with the 180-nm CMOS technology in **Fig. 2(c)**, where "*R*" represents the HfO₂-based RRAM cell. This device consists of a top metal electrode, an HfO₂ resistive switching layer, and a bottom metal electrode. Its conductance state is modulated by controlling the bit line (BL) voltage $V_{BL}$ and word line (WL) voltage $V_{WL}$. When an external voltage is applied between the electrodes, oxygen vacancies in the $HfO_2$ layer are activated, leading to the formation or rupture of conductive filaments and enabling reversible switching between the high-resistance state (HRS) and low-resistance state (LRS). This bipolar switching behavior is clearly demonstrated in **Fig. 2(d)**, which shows the current-voltage (I-V) characteristics of the RRAM cell. **Fig. 2(e)** presents the relationship between the cumulative current on the BL and the number of WLs under different $V_{WL}$. At lower $V_{WL}$ values, e.g., 0.6V and 0.7V, the BL current shows a good linear relationship with the number of WLs. However, at higher $V_{WL}$, the cumulative current increases more rapidly, deviating from the linear trend due to the IR drop issue[46]. This observation suggests that appropriately reducing the WL voltage helps preserve the linearity of BL current, thereby improving the consistency and reliability of array's performance.

**Fig. 2(f)** highlights the key mechanism for this tunable p-bit functionality achieved through in-array random bias (IARB). This method leverages a Gaussian RNG to control the number of activated WLs in the bias region, dynamically modulating the distribution of bias current. The standard deviation $\sigma$ of the bias current distribution can be flexibly adjusted for implementing the DSA algorithm. While $\sigma$ is small, the bias current distribution is narrow and sharply peaked; as $\sigma$ increases, the distribution broadens, and randomness is significantly enhanced. This high-tunability bias current control mechanism forms the foundation for the tunable p-bit functionality. Furthermore, as shown in **Fig. 2(g)**, after the comparison process, the probability of a p-bit outputting a value of 1 under different $\sigma$ values exhibits a typical sigmoidal (*S*-shaped) stochastic characteristic as a function of the Multiply-Accumulate (MAC) input. The range and slope of the probability response vary with $\sigma$, enabling deliberate adjustment steepness of the sigmoid curve to optimize the state update process. **Fig.2(h)** demonstrates the real-time fluctuation behavior of a p-bit under $\sigma = 2$, revealing the dynamic relationship between the p-bit output state and the MAC value. When the MAC value is positive, the p-bit output state is predominantly 1, with occasional state transitions, indicating high determinism toward the high state. When the MAC value is zero, the p-bit output oscillates randomly, with nearly equal proportions of 0 and 1, resembling a random number generator with a uniform binary distribution. For negative MAC values, the p-bit output stays at 0 for most cycles, with infrequent transitions to 1.

**B. Translation of molecular docking to maximum weighted clique problem**

Molecular docking is a complex computational problem that is central to structure-based drug design. It involves identifying the binding interactions between pharmacophore points (PPs) on ligands and receptor proteins, which are critical for determining biological or pharmacological interactions. PPs, characterized by properties such as electrostatic characteristics, hydrophobicity, and hydrogen bond donor/acceptor capabilities, play a crucial role in molecular recognition and binding affinity. In this study, we focus on the LolCDE protein complex, a key component of Gram-negative bacteria, which works in conjunction with the periplasmic chaperone LolA to transport lipoproteins from the inner membrane to the outer membrane[47]. This process is essential for bacterial physiology and pathogenicity, as lipoproteins play crucial roles in cell wall synthesis, bacteria-host interactions, and stress signal transmission[48]. Disrupting this transport process provides a promising strategy for developing antibacterial drugs targeting lipoprotein transport and potentially addresses the growing threat of drug-resistant Gram-negative bacteria. A recent structural biology study utilized high-resolution three-dimensional structures to resolve the ternary complex formed by the LolCDE complex with lipoproteins and the LolA protein, which provided an important structural basis for an in-depth understanding of the molecular mechanism of lipoprotein transport[49].

As illustrated in **Fig. 3(a)**, we identified 6 PPs on the lipoprotein and 7 on the LolA-LolCDE protein complex. For the lipoprotein, these points were classified based on their chemical properties into hydrophobic (hp1–hp4) and hydrogen bond acceptor (ha1) categories. Correspondingly, the PPs on the protein complex were categorized as hydrophobic (HP1–HP4) and hydrogen bond donor (HD1). These PPs theoretically give rise to 42 potential ligand-receptor pharmacophore binding pair modes, where each pair involves one PP from the ligand and one from the receptor protein, as depicted by the connecting lines in **Fig. 3(b)**. Subsequently, an undirected, fully connected graph with 42 vertices was constructed, where each potential pharmacophore binding pair corresponds to a vertex. However, not all pharmacophore binding pairs can coexist in the spatial structure. Therefore, we performed a spatial compatibility check by quantifying the spatial relationship between these PPs on their respective 3D structure based on Euclidean distances (Supplementary Figure S2). Briefly, two binding pairs are deemed compatible only if the distance difference between the PPs satisfies specific criteria (see details in Methods). The adjacency matrix shown in **Fig. 3(c)** captures the coexistence of these 42 vertices after the compatibility check, where black squares indicate compatibility (spatial coexistence of two vertices) and white squares representing incompatibility. Finally, a binding interaction graph (*BIG*), denoted as $G_B = (V, E)$ was generated, as shown in **Fig. 3(d)**. In this graph, each edge $E_{ij} \in E$ indicates that the binding pair represented by vertices $V_i$ and $V_j$ are spatially compatible. Additionally, each vertex $V_i$ is assigned a weight $w_i$. It represents the knowledge-based binding potential of the corresponding binding pair derived from the PDBbind dataset[50–52], as shown in **Table I** (the weight of each vertex are detailed in Supplementary Table S1). These weights reflect the relative strength of interactions and the contribution of each binding pair to the overall molecular binding affinity. The formation of "cliques" within the structure of $G_B$ implies a group of pharmacophore binding pairs that are mutually compatible and collectively define a feasible docking conformation. Among these cliques, the one with the maximum weight is referred to as the Maximum Weighted Clique (MWC), which corresponds to the most energetically favorable and stable docking conformation. This clique maximizes the total binding potential energy while satisfying the spatial compatibility constraints.

To date, the molecular docking problem has been transformed into a MWCP, enabling us to shift focus from the complex, high-dimensional conformational space of ligand-receptor interactions to a well-defined mathematical problem in graph theory. To solve this efficiently, we further reformulate the MWCP into a Quadratic Unconstrained Binary Optimization (QUBO) problem[53], expressed as:

$$F(X) = \sum_{i=1}^{N} w_i x_i + \frac{1}{2} \sum_{i,j}^{N} J_{ij(G_B)} x_i x_j. \tag{3}$$

where the first term represents the total weight of the vertices included in the clique, while the second term accounts for the connectivity amongst the vertices. Matrix $J_{ij(G_B)}$ represents the interconnectivity of vertex $V_i$ and vertex $V_j$ in the BIG. Specifically, $J_{ij(G_B)}$ =1, if the two vertices are connected by an edge, the $J_{ij(G_B)}$ = 0 otherwise.

The maximization of this quadratic function returns a subset of vertices that form a clique while maximizing the total weight. To align with the principle of p-computing and the architecture of our p-computer, we refined the problem formulation. Eq. (3) was reformulated to represent the maximization of the sum of vertex weights in an energy minimization form and penalty terms were introduced to enforce connectivity constraints within the $G_B$:

$$E(X) = -A \cdot \sum_{i=1}^{N} w_i x_i + \frac{P}{2} \cdot \sum_{i,j}^{N} J_{ij(\overline{G_B})} x_i x_j . \tag{4}$$

where the first term minimizes the energy expression to maximize the total weight of selected vertices, while the second term introduces a penalty factor $P$ for pairs of adjacent vertices in the self-complementary graph $\overline{G_B}$ shown as in **Fig. 3(e)**. The relationship of interconnectivity of these two graphs is $J_{ij(\overline{G_B})} = 1 - I_{n \times n} - J_{ij(G_B)}$, where $I_{n \times n}$ is an identity matrix and $n$ is the number of vertices in $G_B$.

Here, the parameter $A$ function as a balance control factor to regulate the influence of the weight term on the total energy. A larger value of $A$ causes the weight term to dominate the energy minimization process; however, this dominance can weaken spatial constraint conditions, potentially resulting in solutions that fail to conform to clique structures. Conversely, increasing $P$ strengthens the spatial constraints among variables in BIG, ensuring that the candidate solutions satisfy the compatibility requirements. However, an excessively large $P$ significantly increases the hardware overhead when mapping $J_{ij}$ matrix to the RRAM chip. To address this issue, in the hardware implementation, we set $A$ and $P$ to 10 and 18 respectively to strike a balance between these two terms. This configuration effectively guides our hardware-based p-computing system toward efficient and accurate energy optimization within the solution space.

**Table I.** Knowledge-based pharmacophore potential derived from the PDBbind dataset. HD: hydrogen-bond donor; HA: hydrogen-bond acceptor; HP: hydrophobic.

| Ligand/Protein | HA | HD | HP |
|:---:|:---:|:---:|:---:|
| **ha** | 0.5478 | 0.6686 | 0.2317 |
| **hd** | 0.6686 | 0.5244 | 0.1453 |
| **hp** | 0.2317 | 0.1453 | 0.0504 |

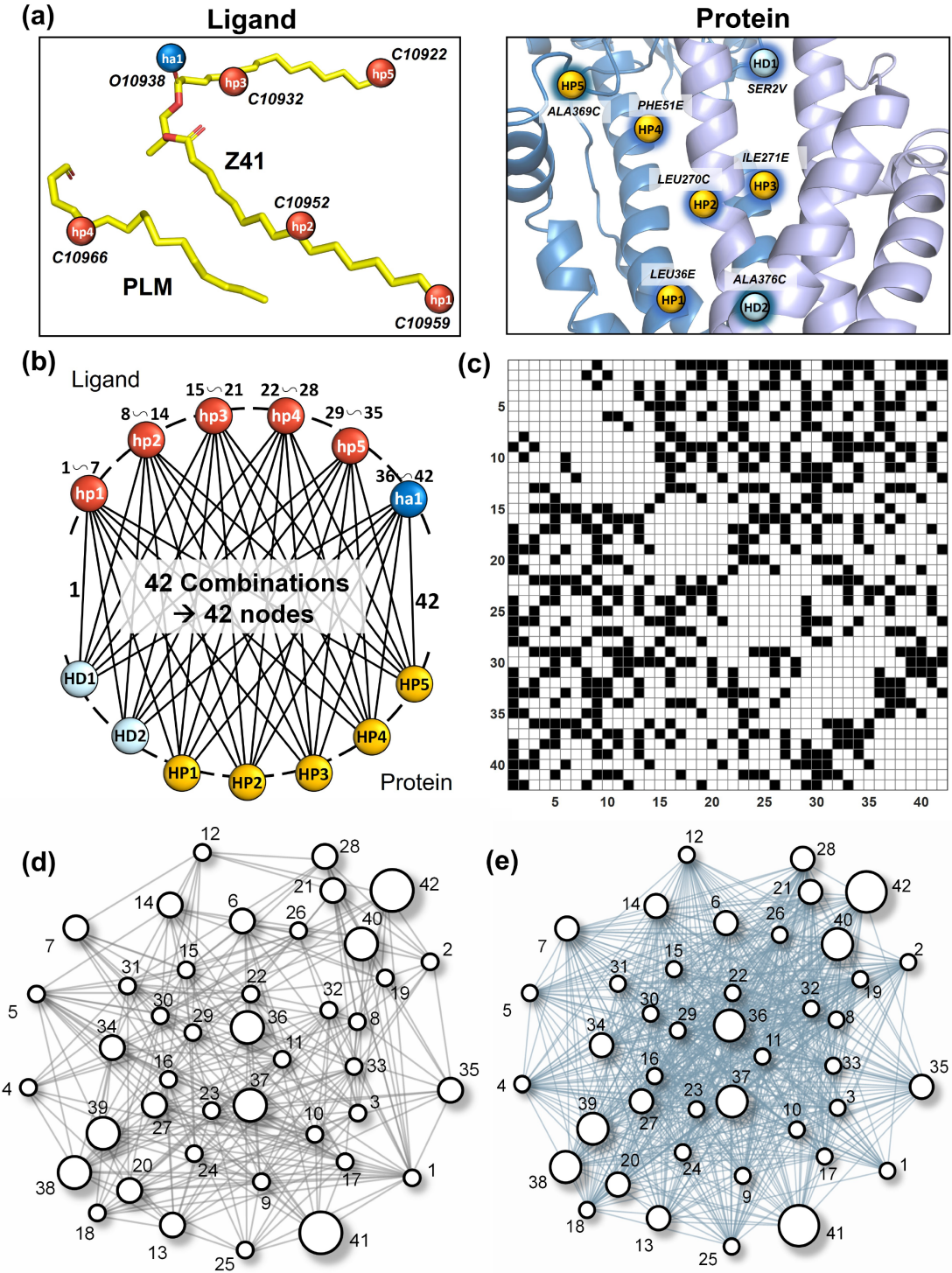

**Fig. 3** (a) Identified 6 PPs on the lipoprotein (left) with the Protein Data Bank (PDB) ID of 7ARM: C10959 (hp1), C10952 (hp2), C10932 (hp3), C10966 (hp4), C10922 (hp5), and O10938 (ha1), located on the PLM and Z41 branches. Identified 7 PPs on the LolCDE-LolA complex (right): HP1 (C3214, residue LEU36E), HP2 (C2057, LEU270C), HP3 (C5034, ILE271E), HP4 (C3313, PHE51E), HP5 (C2758, ALA369C), HD1 (O9421, SER2V), and HD2 (O2805, ALA376C). (b) Pharmacophore binding pairs between the ligand and the protein. Each line represents a potential pharmacophore binding pair and is sequentially named, starting from hp1-HP1 (node 1),…, hp1-HD2 (node 7) to ha1-HD2 (node 42). (c) Adjacency matrix after the spatial compatible check. (d) $G_B$ of this docking problem and its (e) self-complementary graph, where vertices with different weights are shown in different sizes.

## C. Experimental demonstration of a 42-node molecular docking

**Identifying MWCs with the p-computer under fixed stochasticity levels**

To systematically evaluate the performance of our RRAM-based p-computer in real molecular docking scenarios, framed as a MWCP, we modulated the stochasticity of the system by varying the standard deviation $\sigma$ of the Gaussian RNG. Fig. 4 illustrates the energy fluctuation profiles during iterations, the solution distributions and the final predicted MWC across 100 independent experimental trials for each $\sigma$ setting.

Under conditions of constant stochasticity, the operating mechanism of the p-computer is governed by Boltzmann's Law:

$$P(X) = \frac{\exp\left(-\frac{E(X)}{T}\right)}{\sum_{i,j} \exp\left(-\frac{E(X)}{T}\right)} \quad (5)$$

where $T$, referred to as the pseudo-temperature parameter in the context of p-computing, characterizes the system's stochasticity. Consequently, the MWC corresponding to the ground state is expected to exhibit the highest probability. Following this principle, our p-computer identifies the most frequently occupied p-bit state configuration in each trial, represents it as a subgraph, and designates it as the trial's output.

**Fig. 4(a)** shows that at the highest level of stochasticity ($\sigma = 2.0$), the system exhibited pronounced energy fluctuations, indicative of extensive and thorough exploration of the solution space. However, despite transiently reaching the theoretical ground state $E = -8.702$ at various points during the iterative process, which corresponds to an unidentified clique with a cumulative weight of 0.8702, the system ultimately failed to consistently occupy this optimal solution due to insufficient stability. Instead, the system predominantly produced lower-weight configurations, represented by (11, 17, 23, 41) and (5, 17, 23, 41), with cumulative weights of 0.8198 and occurrence frequencies of 32% and 29%, respectively.

Reducing $\sigma$ to 1.5 (**Fig. 4b**) and 1.0 (**Fig. 4c**) mitigated the energy fluctuations, signaling a transition from broad exploration to localized oscillations around lower-energy configurations near $-4$ and $-8$. At these two stochasticity levels, the output subgraphs became more concentrated with the clique (5, 17, 23, 41) consistently emerging as the dominant solution in 36% and 37% of trials, respectively. However, despite the improved focus on low-energy configurations, like the clique with $E = -8.702$ ranked as the second most frequently observed output at $\sigma = 1.0$, the system still failed to reliably identify this optimal solution. It suggests the system's insufficient convergence capability under these stochasticity levels.

As $\sigma$ was further reduced to 0.5 (**Fig. 4d**), the system achieved a critical balance between exploration and identification. The considerably reduced stochasticity led to more confined energy fluctuations, shown as a distinct energy stratification at $E = -5$ and $-8$. Under this condition, in nearly one-third of the 100 trials, the system occupied the ground state for the longest duration during the iterations, thus successfully identifying the theoretically optimal solution for the first time. The MWC (1, 9, 17, 25, 41) with a cumulative weight of 0.8702 emerged as the most frequent output clique, accounting for 35% of the trials. On the other hand, other candidate configurations, such as cliques (1, 26, 42) and (1, 21, 40), persisted, appearing in 15% and 12% of the outcomes, respectively. These lower-weight cliques represent local minima within the energy landscape, where the system occasionally settles into during the evolvement process. Finally, at the lowest stochasticity level ($\sigma = 0.2$, **Fig. 4e**), the system exhibited minimal energy fluctuations. At this point, in addition to the Gaussian RNG, the contribution of device-to-device variations within the RRAM array becomes more prominent, but still in a low range. While this setting led to a more concentrated solution distribution, the system predominantly identified a non-optimal clique, (1, 26, 42), in 37% of trials, with the ground state occurring in only 26%. This outcome reveals a critical limitation of excessively low stochasticity, where restricted exploration leads the system to local optima, compromising its ability to achieve the ground state.

**Fig. 5(a)** summarizes the success probabilities of our RRAM-based p-computer in identifying the MWC, providing a comparative analysis of experimental and simulation results across these five stochasticity levels. Both datasets present a consistent trend: the success probability initially rises with increasing stochasticity, reaching an optimal performance at $\sigma = 0.5$, before declining as stochasticity further be enhanced. The observed differences between experimental and simulation results are primarily attributed to device-to-device variations inherent in the physical RRAM hardware, which are not fully captured in the simulations. Nevertheless, the similar convergence of trends and improvements of success probabilities demonstrates the system's optimal capability in balancing the solution space exploration and MWC identification at $\sigma = 0.5$.

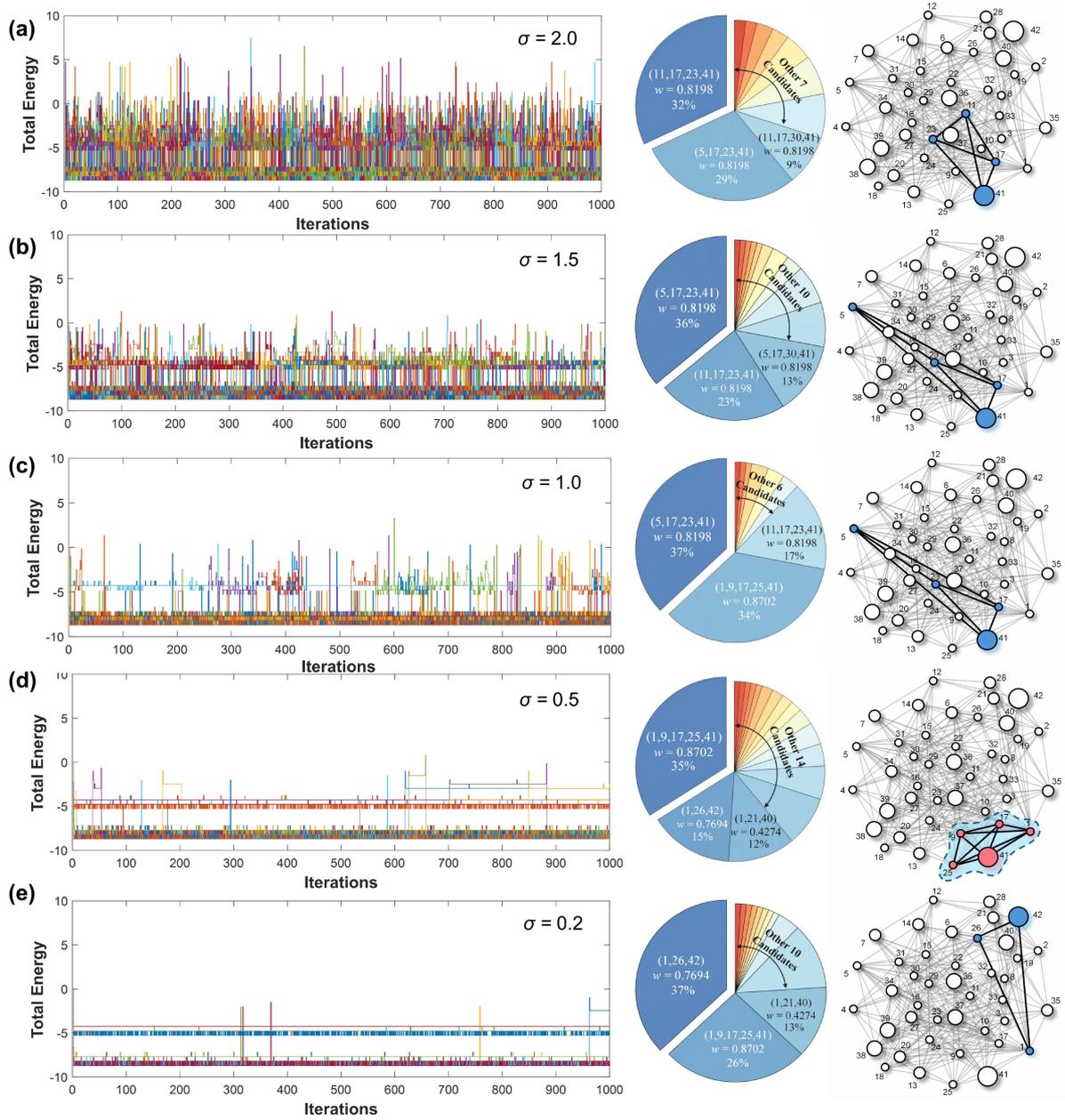

**Fig. 4** Energy dynamics during iterations (left), solution distributions (middle), and predicted MWCs (right) for varying levels of stochasticity: (a) $\sigma = 2.0$, (b) 1.5, (c) 1.0 (d) 0.5 and (e) 0.2, based on 100 independent trials per condition. Detailed experimental output subgraphs corresponding to these five stochasticity levels are provided in Supplementary Tables S2, S3, S4, S5, and S6, respectively.

**Identifying MWCs with the p-computer under DSA**

We then introduced a novel DSA strategy tailored for this p-computer to further improve performance of system. In this approach, the stochasticity of all p-bits is globally modulated from their respective initial $\sigma$ values to $\sigma = 0$. **Fig. 5(b)** compares the experimental success probabilities of identifying the MWC in 100 trials with the application of DSA approach against previous constant stochasticity settings. Without DSA, the success probability peaks at 35% for $\sigma = 0.5$. However, at lower stochasticity ($\sigma = 0.2$), restricted exploration leads to fluctuations among local energy minimums, while at higher stochasticity ($\sigma = 1.5, 2.0$), excessive noise inhibits the system's ability to stabilize near optimal solutions. In contrast, the DSA strategy significantly improves success probability, achieving a peak at 72% by dynamically reducing $\sigma$ from 0.5 to 0. This comparison underscores the importance of stochastic modulation in the p-computer, which enables the system to adaptively transition from global exploration to localized convergence. However, it should be noted that it does not always guarantee the improvement of MWC identification across all stochasticity levels, such as $\sigma = 1.0$. Factors such as device to device variations and inappropriate DSA schedule for a specific $\sigma$ can limit the performance of this strategy.

A complete energy evolution of the system under the DSA application in one 1000-iteration process is shown in **Fig.5(c)**. During the initial phase, the system exhibits significant energy fluctuations. This high level of stochasticity enables the system to effectively escape local minima and conduct a comprehensive exploration of the solution space. As DSA progresses and the stochasticity level gradually diminishes, the system transitions toward lower energy states, exhibiting increasingly stable oscillatory behavior. In the final stage of annealing, as $\sigma$ approaches 0, the system's energy stabilizes near the ground state ($E = -8.702$). However, due to the inherent device-to-device variations in the RRAM hardware, the system retains a small degree of randomness even when the Gaussian RNG's $\sigma$ is near 0. As shown in the left inset, this hardware-induced perturbation results in energy oscillations between global ground state and a sub-optimal state ($E = -8.198$), with the system exhibiting stronger preference for the ground state. **Fig. 5(d)** presents the complete output subgraph distributions of the detailed 100 experimental trials conducted with DSA. Among the identified solutions, the clique (1,9,17,25,41), which corresponds to the theoretical ground state with a cumulative weight of 0.8702, emerged as the dominant outcome, achieving a success probability of 72%. In comparison, other candidate solutions, such as the cliques (1, 21, 40) with a weight of 0.4274 and (26, 29, 42) with a weight of 0.7694, occurred far less frequently, with success probabilities of 8% and 4%, respectively. This discrepancy highlights the effectiveness of the DSA approach in guiding the system toward achieving the global optimum.

The MWC determined through this process is visualized in **Fig. 5(e)**. The resulting isolated subgraph highlights the vertices comprising the optimal solution, which directly corresponds to the matched PP binding pairs in this 42-node molecular docking problem. Specifically, vertices 1, 9, 17, 25, and 41 represent the binding PP pairs hp1-HP1, hp2-HP2, hp3-HD3, hp4-HP4 and ha1-HD1, respectively. **Fig. 5(f)** provides the three-dimensional visualization of the predicted docking configuration between the lipoprotein and the LolCDE-LolA complex reconstructed from the identified MWC. This configuration includes four pairs of hydrophobic interactions and one hydrogen bond, collectively representing the most stable docking conformation with the highest binding affinity. Notably, the predicted binding mode closely aligns with the protein-ligand interactions analyzed using the Protein-Ligand Interaction Profiler (Supplementary Figure S3)[54,55], further validating the accuracy and reliability of our p-computer.

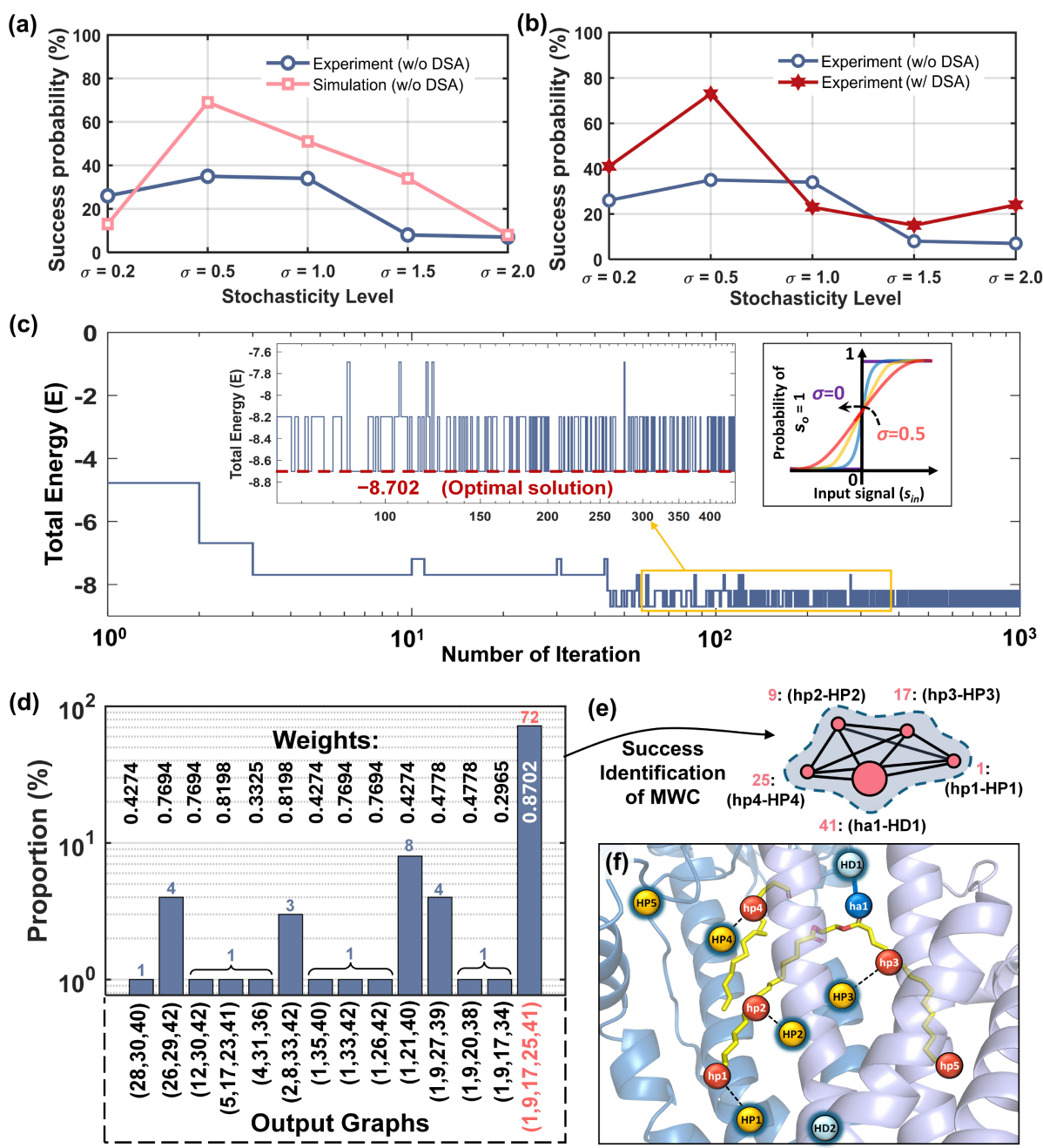

**Fig. 5** Performance evaluation of the p-computer without and with the application of DSA in solving the 42-node molecular docking problem of the lipoprotein and the LolCDE-LolA complex. (a) Success probability comparison between experiments and simulations under constant stochasticity levels. (b) Experimental success probability comparison with and without the application of DSA. (c) Dynamic evolution of the system's energy during a complete 1000-iteration DSA process. Inset (left): detailed energy evolution at iterations 60~440. Inset (right): applied DSA schedule of σ decreased from 0.5 to 0. (d) Distribution of output solutions from 100 effective experimental trials. Detailed experimental output subgraphs are provided in Supplementary Table S7. (e) Graphical representation of the identified optimal MWC, where the vertices and its corresponding exact binding PP pairs are highlighted. (f) Cartoon representation of the predicted docking configuration between the ligand and the receptor protein.

## Conclusion and Discussion

This work presents a groundbreaking approach to solving molecular docking challenges by leveraging the novel hardware and algorithmic capabilities of a p-computer based on tunable artificial p-bits. Through a comprehensive hardware prototype integrating 180 nm CMOS and $HfO_2$-based RRAM technology, we successfully demonstrated the feasibility and effectiveness of the p-computer in addressing the challenging MWCP reformulation of molecular docking. Notably, our system achieved a success probability of approximately 72% in resolving a 42-node MWCP derived from the LolA-LolCDE-lipoprotein complex, outperforming existing methods in terms of both problem scale and success probability. The comparative analysis provided in Table II highlights the distinct advantages of our RRAM-based p-computer over alternative graph problem-based molecular docking solvers, including GBS systems[38,39], DC-QAOA[41] and a programmable photonic processor[40]. Unlike these quantum-based methods, which are typically limited by hardware scalability and the necessity for post-processing algorithms, our p-computer delivers a seamless, one-step solution without the need for additional computational corrections. This superior performance is primarily attributed to the high-quality solutions generated by the p-computer. By an appropriate design of the penalty parameter, the p-computer can predominantly explore candidate subgraphs that satisfy the clique property. As a result, for these output subgraphs, the post-processing algorithm provides negligible performance enhancement (results are provided in Supplementary Tables S8 and S9), as the solutions cannot be further optimized to achieve higher weights. Another particularly notable limitation of quantum photonic platforms, such as GBS systems, lies in their dependence on large-scale experimental setups that frequently occupy the space of an entire laboratory, rendering them impractical for real-world deployment. In contrast, our p-computer, implemented on a compact PCB board with a 6.5 × 4 mm² integrated RRAM chip, offers exceptional practicality for real-world molecular docking scenarios. Furthermore, the architectural design of the p-computer incorporates a CIM scheme that effectively decouples of stochasticity from intrinsic device characteristics. This innovative design not only reduces the technical barriers to implementing p-bits and p-computers in hardware but also paves the way for the widespread adoption of p-computing across a diverse range of optimization tasks.

The implications of this work also extend well beyond molecular docking, positioning p-computing as a transformative and versatile technology for addressing complex combinatorial optimization challenges across diverse scientific domains. By demonstrating exceptional scalability, hardware efficiency, and algorithmic elegance in practical hardware implementation, this study represents the first successful application of p-computing in the molecular docking-based computational biology area. It provides a solid foundation and a template for future advancements in drug discovery and other related fields.

**Table II.** Comparison of state-of-art graph problem-based molecular docking solvers

| Solvers | Implementation Modality | Specific Problem | Problem Scale | Success Probability Reported | Postprocessing Algorithms Necessary? | Experimental Equipment Footprint | Scalability |
|---|---|---|---|---|---|---|---|
| RRAM-based p-computer (this work) | Experimental (Integrated Electronics) | MWCP | 42 nodes | 72% | No | 6.5×4 mm²[a] | Good |
| GBS + Postprocessing[38] | Simulated | MWCP | 24 nodes | ~70% | Yes | — | Poor |
| GBS ("Abacus") + Postprocessing[39] | Experimental (Free-Space Optics) | MWCP | 28 nodes | ~32% | Yes | 2m²[b] | Poor |
| DC-QAOA[41] | Simulated | MWCP | 12 nodes | ~51% | No | — | Poor |
| Photonic Processor[40] | Experimental (Integrated Photonics) | MCP | 9 nodes | — | No | 50 mm²[b] | Poor |

[a] Only chip size is considered. This p-computer is implemented on a printed circuit board (PCB) that integrates a RRAM-CMOS chip.
[b] Estimated value.

## Methods

### Fabrication of BEOL RRAM device integrating on 180nm CMOS technology

The RRAM device, comprising TiN/HfO$_x$/Ti/TiN stacked layers, was integrated into a CMOS array to develop a large-scale 1Mb RRAM p-computer macro. During the BEOL processing, a TiN metal layer, serving as the bottom electrode (BE), was deposited using a physical vapor deposition (PVD) system on the top of the tungsten plug connected to the drain (D) node of the transistor. Subsequently, the BE was patterned and planarized through chemical mechanical polishing (CMP). A 6 nm-thick HfO$_x$ resistive switching layer was then deposited using an atomic layer deposition (ALD) system (ASM, Polygon 8200) at a temperature of 300°C. To further enhance the device structure, a 10 nm-thick Ti buffer layer was deposited over the HfO$_x$ layer using the PVD system. Without breaking the vacuum, a 40 nm-thick TiN metal layer, functioning as the top electrode (TE), was deposited using the same PVD system. The memory stacked layers were subsequently patterned via standard lithography and passivated with low-temperature oxide (LTO). Finally, the fabricated devices were then annealing at 400°C for 5 minutes in the N$_2$ environment.

### Architecture of 1T1R RRAM p-computer and details for algorithm mapping

The p-computer chip, based on 1T1R binary RRAM technology, was fabricated using a 180 nm CMOS process with BEOL HfO2 RRAM integration. This chip features a 1152 × 1024 1T1R RRAM array, paired with 512 current-mode sense amplifiers (CSA) to compare the MAC outputs of every two bitlines (BLs). To solve the MWCP reformulation of the molecular docking problem, the weights and biases are encoded in paired neighboring BLs using paired

RRAM devices. For a positive value, the left RRAM device is programmed to the forming state, while the right RRAM device remains in the non-forming state. Conversely, a negative value is encoded by programming the right RRAM device to the forming state, with the left RRAM device remaining in the non-forming state. To prepare the in-array bias region, 36 paired RRAM cells are programmed to represent both positive and negative biases, with 18 cells allocated for positive bias and 18 cells for negative bias. The Gaussian RNG is used to determine the bias value for each iteration. The two MAC values of paired neighboring bitlines (BLs) are compared using either a CSA or an analog-to-digital converter (ADC) to determine the results during each read step. A microcontroller unit (MCU)-based PCB test board is utilized to manage the data flow.

**MCU System Board**

The MCU system board is designed with the CIM chip at its core, serving as the primary control and interface platform. The MCU is responsible for providing digital signal control, including critical inputs such as write addresses and input positions. These signals are first processed through a level shifter, which adjusts them to the appropriate operating voltage before transmitting them to the CIM chip via jumpers. Similarly, output signals from the CIM chip are routed back through the level shifter before being received and processed by the MCU. To facilitate stable and precise operation, a DAC (Digital-to-Analog Converter) is integrated on the right side of the system board. This DAC supplies various analog voltages, including operating voltage (VDD), write voltage, read voltage, calibration voltage, and PAD power supply voltage, all of which are carefully regulated to ensure reliable system performance.

**Geometric Compatibility Check of Binding Pairs**

Geometric compatibility check is an essential step in constructing the BIG for the target molecular docking problem. This rigorous filtering process can minimize the inclusion of physically implausible or sterically unfavorable configurations, thereby reducing the solution space and enhancing the accuracy and reliability of downstream docking predictions performed by the p-computer.

The geometric compatibility of pharmacophore pairs is assessed based on their Euclidean distances. Specifically, for two selected pharmacophore binding pairs, ($L_1$, $P_1$) and ($L_2$, $P_2$), where $D_1$ represents the distance between PPs $L_1$ and $L_2$ on the ligand and $D_2$ represents the distance between PPs $P_1$ and $P_2$ on the receptor protein, the binding pairs are consider geometrically compatible if and only if the absolute difference between $D_1$ and $D_2$ satisfies the condition $|D_1 - D_2| \leq \tau + 2\varepsilon$. Here, $\tau$ denotes the flexibility constant, and $\varepsilon$ signifies the interaction distance tolerance. For the setting of parameter values of $\tau$ and $\varepsilon$ used in the compatibility check, please refer to Ref. [38] and [41] for more details. In this study, $\tau$ and $\varepsilon$ is set to 0.1 Å and 3.3 Å, respectively.

**Post-Processing in Graph-Based Molecular Docking**

In solving graph problems-based molecular docking, inherent losses and noise in GBS often result in output subgraphs that fail to satisfy the clique property. To address these limitations, previous GBS-related studies[38,39] have implemented a post-processing algorithm to optimize these suboptimal subgraphs to enhance the identification of the MWC and increase the success probability of finding the optimal docking configuration. The workflow of this hybrid GBS + postprocessing method begins with GBS performing global exploration to generate subgraphs as initial solutions and then the post-processing algorithm takes these subgraphs as input seeds to enhance solution quality through local optimization. To ensure a fair comparison of the performance between our p-computer and the GBS + post-processing method, we adopted the same post-processing algorithm described in the GBS-related works, implemented using the Python toolkit provided by the Strawberry Fields platform[44].

Specifically, in each trial, the experimental output subgraph from the p-computer, defined as the most frequently visited state configuration during the iterative process, was used as the input seed for the post-processing algorithm. The post-processing algorithm consists of two key steps: shrinking and local search. During the shrinking step, the algorithm iteratively reduces the size of the input subgraph until a clique is formed. In each iteration, vertices with the lowest degree relative to the subgraph and the smallest weight are removed based on predefined criteria. This strategy can effectively retain nodes more likely to form the MWC while excluding redundant or noisy nodes. The local search step, based on the dynamic or phased local search algorithms described in Refs. [42] and [43], further optimizes the clique obtained from the shrinking step. This step alternates between two phases: greedy expansion and plateau search. During the greedy expansion phase, the algorithm incrementally adds nodes connected to the current clique with the highest weight until no further additions are possible. In the plateau search phase, the algorithm explores the neighborhood of the current clique to identify structures with potentially higher weights. By alternating between these two phases, the local search algorithm effectively refines the initial solution obtained from the contraction step, approaching the global optimum.

The performance of post-processing to the p-computer's output was evaluated with and without DSA under a stochasticity level of $\sigma = 0.5$, with detailed experimental results provided in Supplementary Tables S8 and S9.

## Acknowledgements

This work at the National University of Singapore was supported by FRC-A-8000194-01-00. G.C. L. would also like to thank the financial support from the National Science and Technology Council (NSTC) under grant number NSTC 112-2112-M-A49-047-MY3, the Co-creation Platform of the Industry-Academia Innovation School, NYCU, under the framework of the National Key Fields Industry-University Cooperation and Skilled Personnel Training Act, and the Advanced Semiconductor Technology Research Center from The Featured Areas Research Center Program within the framework of the Higher Education Sprout Project by the Ministry of Education (MOE) in Taiwan.

## Data availability

The data that support the plots within this paper and the other findings of this study are available from the corresponding author upon reasonable request.

## Code availability

The computer code and problem instances used in this study are available from the corresponding authors upon reasonable request.

## Author contributions

H.Y., H.M., H.T., and L.G. conceptualized the study and proposed the idea of solving molecular docking problems with probabilistic computing. H.Y. developed the molecular docking framework with the help of D.Q. and designed the simulation experiments. H.Y. performed the simulations with technical assistance from F.C. H.M., designed the experimental protocols, constructed the experimental setup, including the fabrication of the RRAM-based chip, with assistance from L.C.-S. and L.C.-M. H.M. conducted experiments based on the developed probabilistic computer hardware. H.Y. and H.M., with input from G.X., H.T., and L.G., analyzed the simulation and experimental data. H.Y. and H.M. drafted the manuscript with input from all other authors. All authors discussed the results and reviewed the manuscript.

## Corresponding authors

Correspondence and requests for materials should be addressed to L.G. (Email: gcliang@nycu.edu.tw) and H. T. (Email: thhou@nycu.edu.tw).

## Competing interests

The authors declare no competing interests.

# Supplementary Materials for

# The First Hardware Demonstration of a Universal Programmable RRAM-based Probabilistic Computer for Molecular Docking

Yihan He[1†], Ming-Chun Hong[2,3†], Qiming Ding[4], Chih-Sheng Lin[2,3], Chih-Ming Lai[3], Chao Fang[1], Xiao Gong[1], Tuo-Hung Hou[2,5*], and Gengchiau Liang[1,5*]

[1]Department of Electrical and Computer Engineering, National University of Singapore, 117583 Singapore
[2]Department of Electrical Engineering and Institute of Electronics, National Yang-Ming Chiao Tung University, Hsinchu, Taiwan
[3]Electronic and Optoelectronic System Research Laboratories, Industrial Technology Research Institute, Hsinchu, Taiwan
[4]Center on Frontiers of Computing Studies, Peking University, Beijing 100871, China
[5]Industry Academia Innovation School, National Yang-Ming Chiao Tung University, Hsinchu, Taiwan

[†]These authors contribute equally;   [*]Email: thhou@nycu.edu.tw; gcliang@nycu.edu.tw

**This PDF file includes:**
Supplementary Figures S1 to S3,
Supplementary Tables S1 to S9.

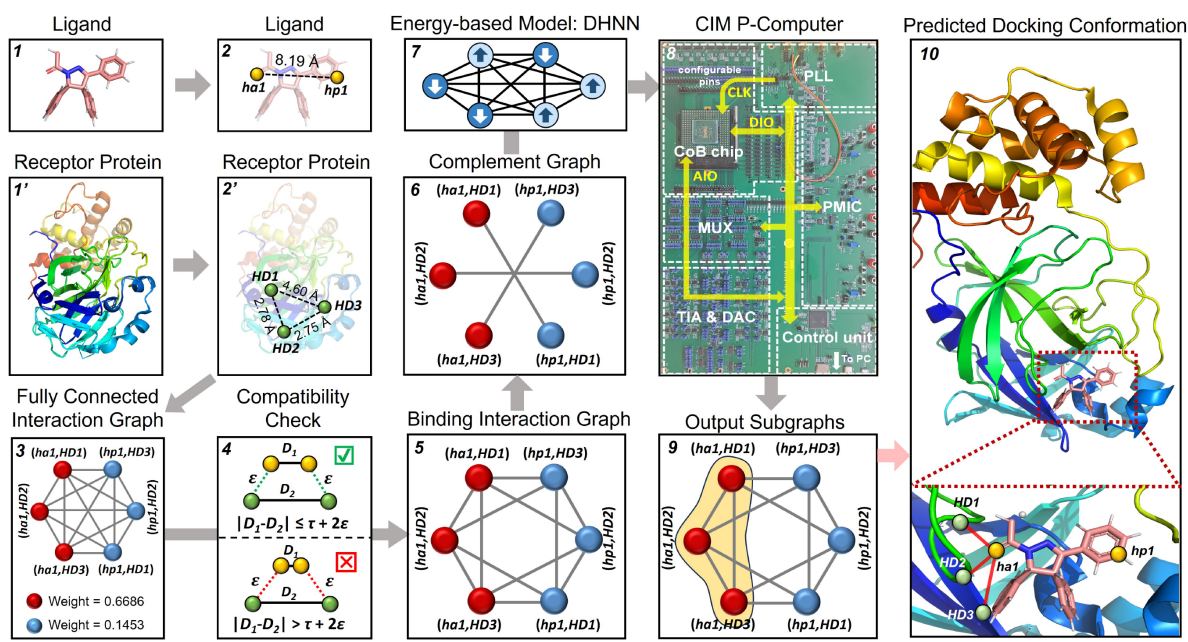

**Supplementary Figure S1** Schematic diagram of the protein-ligand docking process solved under the p-computing framework, using the example of SARS-CoV-2 Mpro (COVID-19) docking with covalent pyrazoline-based inhibitors PM-2-020B (PDB ID: 8SKH). Step (1): Visualization of 3D structures of the protein and receptor protein molecules. Step (2): Identification and distance calculation of pharmacophore points on the ligand and receptor protein. For clearer illustration, we identified 2 and 3 pharmacophore points on the ligand and receptor protein, respectively. Step (3): Construction of the fully connected graph based on all combinations of pharmacophore pairs. Step (4): Compatibility check of the connections between vertices based on spatial distances. Step (5): Evolution into a BIG after the compatibility check. Step (6): Generation of the self-complementary graph of the BIG. Step (7): Mapping of the self-complementary graph through the DHNN. Step (8): Development of a p-computer built upon p-bits. Step (9): Identification of the MWC from the p-computer's output, as shown by the three red vertices covered by the yellow shadow. Step (10): Restoration of the final docking conformation from the solution of the MWCP and its enlarged display.

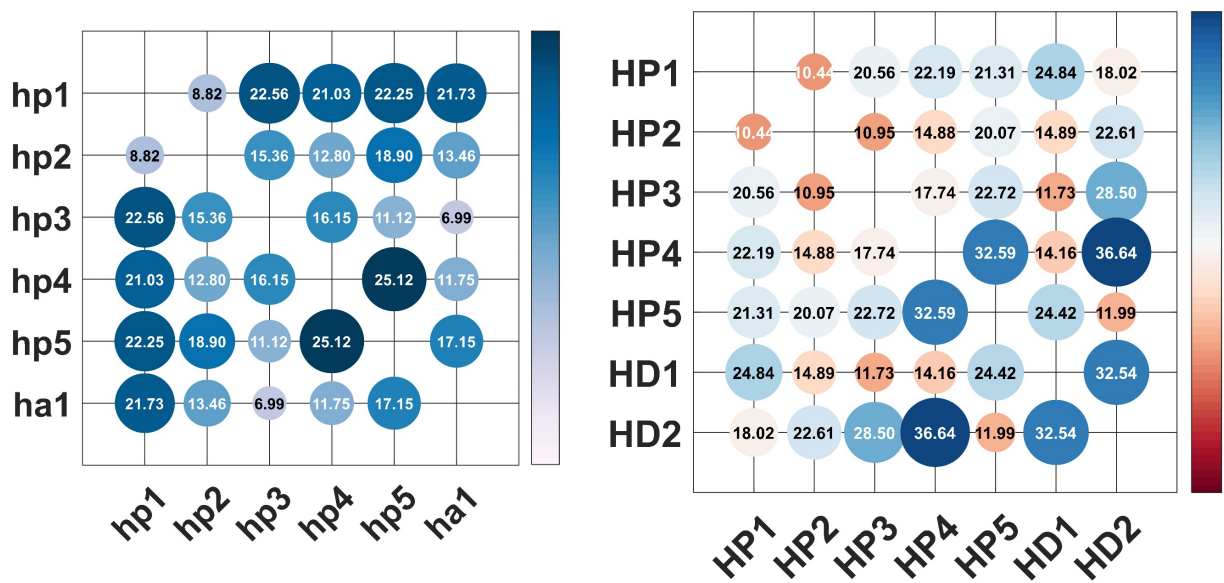

**Supplementary Figure S2** Pairwise distance matrices of the PPs on the lipoprotein (left) and (e) the LolCDE-LolA complex (right), which are calculated based on the known three-dimensional structure in PDB file.

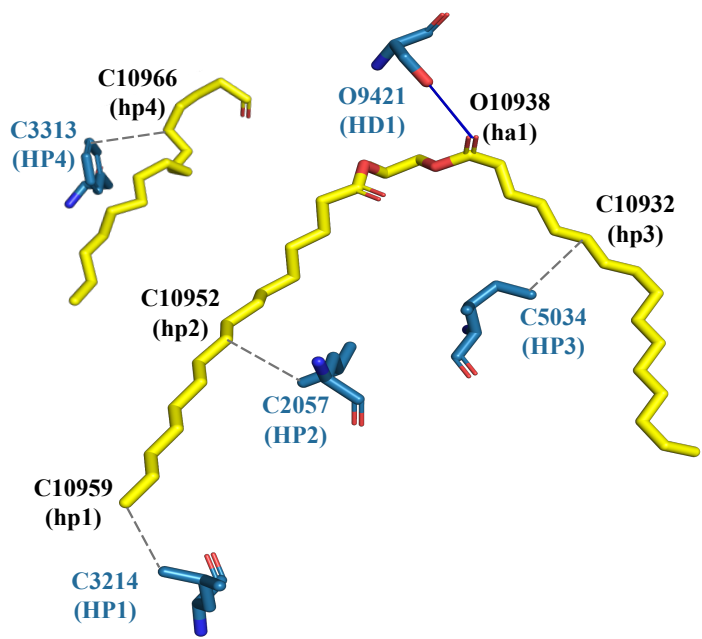

**Supplementary Figure S3** Visualization of the binding sites of the lipoprotein and the LolCDE-LolA complex (PDB ID: 7ARM) analyzed by the Protein-Ligand Interaction Profiler. The yellow structure represents the lipoprotein, while the blue structures correspond to the interacting amino acid residues on the receptor protein. Interaction types are depicted with dashed grey lines for hydrophobic interactions and a solid blue line for the hydrogen bond.

**Supplementary Table S1.** The weight of each vertex in the 42-node $G_{BIG}$.

| Vertices | Weights |
|---|---|
| 1(hp1-HP1), 2 (hp1-HP2), 3(hp1-HP3), 4 (hp1-HP4), 5(hp1-HP5) 8(hp2-HP1), 9 (hp2-HP2), 10(hp2-HP3), 11(hp2-HP4), 12(hp2-HP5) 15(hp3-HP1), 16(hp3-HP2), 17(hp3-HP3), 18(hp3-HP4), 19(hp3-HP5) 22(hp4-HP1), 23(hp4-HP2), 24(hp4-HP3), 25(hp4-HP4), 26(hp4-HP5) 29(hp5-HP1), 30(hp5-HP2), 31(hp5-HP3), 32(hp5-HP4), 33(hp5-HP5) | 0.0504 |
| 6 (hp1-HD1), 7 (hp1-HD2), 13(hp2-HD1), 14(hp2-HD2), 20(hp3-HD1) 21(hp3-HD2), 27(hp4-HD1), 28(hp4-HD2), 34(hp5-HD1), 35(hp5-HD2) | 0.1453 |
| 36(ha1-HP1), 37(ha1-HP2), 38(ha1-HP3), 39(ha1-HP4), 40(ha1-HP5) | 0.2317 |
| 41(ha1-HD1), 42(ha1-HD2) | 0.6686 |

**Supplementary Table S2** Detailed results of 100 effective experimental trials corresponding to Fig. 4(a), conducted under a constant $\sigma = 2.0$. Each row represents a distinct output configuration, including its 42-bit binary representation, equivalent decimal value, associated output graphs, and the corresponding docked PP pairs. The table also includes the proportion, and total weights of each output graph.

| No. | Exp. Output in Binary format (42-bit) | Decimal | Output Graphs | Corresponding Docked PP pairs | Proportion (%) | Weights |
|---|---|---|---|---|---|---|
| 1 | '000000000000010000000000000000010000000001' | 1073745921 | (12, 30, 42) | (hp2-HP5) (hp5-HP2) (ha1-HD2) | 3 | 0.7694 |
| 2 | '000000000000010000000000000010000000000001' | 1073750017 | (12, 29, 42) | (hp2-HP5) (hp5-HP1) (ha1-HD2) | 2 | 0.7694 |
| 3 | '000000000010000010000000000001000000000010' | 2181042178 | (11, 17, 30, 41) | (hp2-HP4) (hp3-HP3) (hp5-HP2) (ha1-HD1) | 9 | 0.8198 |
| 4 | '000000000010000010000001000000000000000010' | 2181562370 | (11, 17, 23, 41) | (hp2-HP4) (hp3-HP3) (hp4-HP2) (ha1-HD1) | 32 | 0.8198 |
| 5 | '000010000000000010000000000001000000000010' | 137472512002 | (5, 17, 30, 41) | (hp1-HP5) (hp3-HP3) (hp5-HP2) (ha1-HD1) | 8 | 0.8198 |
| 6 | '000010000000000010000001000000000000000010' | 137473032194 | (5, 17, 23, 41) | (hp1-HP5) (hp3-HP3) (hp4-HP2) (ha1-HD1) | 29 | 0.8198 |
| 7 | '010000010000000000000000000001000000000010' | 1116691497473 | (2, 8, 33, 42) | (hp1-HP2) (hp2-HP1) (hp5-HP5) (ha1-HD2) | 3 | 0.8198 |
| 8 | '100000000000000000000001000000000000000001' | 2199023321089 | (1, 26, 42) | (hp1-HP1) (hp4-HP5) (ha1-HD2) | 2 | 0.7694 |
| 9 | '100000000000000010000010000000000000000010' | 2199056941058 | (1, 17, 25, 41) | (hp1-HP1) (hp3-HP3) (hp4-HP4) (ha1-HD1) | 5 | 0.8198 |
| 10 | '100000001000000010000010000000000000000010' | 2207646875650 | (1, 9, 17, 25, 41) | (hp1-HP1) (hp2-HP2) (hp3-HP3) (hp4-HP4) (ha1-HD1) | 7 | 0.8702 |

**Supplementary Table S3** Detailed results of 100 effective experimental trials corresponding to Fig. 4(b), conducted under a constant $\sigma = 1.5$.

| No. | Exp. Output in Binary format (42-bit) | Decimal | Output Graphs | Corresponding Docked PP pairs | Proportion (%) | Weights |
|---|---|---|---|---|---|---|
| 1 | '000000000000000000000000010010000000000001' | 73729 | (26, 29, 42) | (hp4-HP5) (hp5-HP1) (ha1-HD2) | 1 | 0.7694 |
| 2 | '000000000001000000000000000010000000000001' | 1073750017 | (12, 29, 42) | (hp2-HP5) (hp5-HP1) (ha1-HD2) | 1 | 0.7694 |
| 3 | '000000000010000010000000000001000000000010' | 2181042178 | (11, 17, 30, 41) | (hp2-HP4) (hp3-HP3) (hp5-HP2) (ha1-HD1) | 7 | 0.8198 |
| 4 | '000000000010000010000000010000000000100000' | 2181070880 | (11, 17, 27, 37) | (hp2-HP4) (hp3-HP3) (hp4-HD1) (ha1-HP2) | 1 | 0.4778 |
| 5 | '000000000010000010000100000000000000000010' | 2181562370 | (11, 17, 23, 41) | (hp2-HP4) (hp3-HP3) (hp4-HP2) (ha1-HD1) | 23 | 0.8198 |
| 6 | '000010000000000000000000000010000000000100' | 68719484932 | (6, 29, 40) | (hp1-HD1) (hp5-HP1) (ha1-HP5) | 1 | 0.4274 |
| 7 | '000010000000000000000000000001000000000010' | 137472512002 | (5, 17, 30, 41) | (hp1-HP5) (hp3-HP3) (hp5-HP2) (ha1-HD1) | 13 | 0.8198 |
| 8 | '000010000000000010000100000000000000000010' | 137473032194 | (5, 17, 23, 41) | (hp1-HP5) (hp3-HP3) (hp4-HP2) (ha1-HD1) | 36 | 0.8198 |
| 9 | '000010000000000010001000000000000100100000' | 137473556768 | (5, 17, 22, 34, 37) | (hp1-HP5) (hp3-HP3) (hp4-HP1) (hp5-HD1) (ha1-HP2) | 3 | 0.5282 |
| 10 | '010000010000000000000000000001000000000001' | 1116691497473 | (2, 8, 33, 42) | (hp1-HP2) (hp2-HP1) (hp5-HP5) (ha1-HD2) | 3 | 0.8198 |
| 11 | '100000000000000000000010000000000000000001' | 2199023321089 | (1, 26, 42) | (hp1-HP1) (hp4-HP5) (ha1-HD2) | 1 | 0.7694 |
| 12 | '100000000000000010000100000000000000000010' | 2199056941058 | (1, 17, 25, 41) | (hp1-HP1) (hp3-HP3) (hp4-HP4) (ha1-HD1) | 2 | 0.8198 |
| 13 | '100000001000000010000100000000000000000010' | 2207646875650 | (1, 9, 17, 25, 41) | (hp1-HP1) (hp2-HP2) (hp3-HP3) (hp4-HP4) (ha1-HD1) | 8 | 0.8702 |

**Supplementary Table S4** Detailed results of 100 effective experimental trials corresponding to Fig. 4(c), conducted under a constant $\sigma = 1.0$.

| No. | Exp. Output in Binary format (42-bit) | Decimal | Output Graphs | Corresponding Docked PP pairs | Proportion (%) | Weights |
|---|---|---|---|---|---|---|
| 1 | '000000000000000000000000000101000000000100' | 20484 | (28, 30, 40) | (hp4-HD2) (hp5-HP2) (ha1-HP5) | 1 | 0.4274 |
| 2 | '000000000010000010000000000001000000000010' | 2181042178 | (11, 17, 30, 41) | (hp2-HP4) (hp3-HP3) (hp5-HP2) (ha1-HD1) | 3 | 0.8198 |
| 3 | '000000000010000010000100000000000000000010' | 2181562370 | (11, 17, 23, 41) | (hp2-HP4) (hp3-HP3) (hp4-HP2) (ha1-HD1) | 17 | 0.8198 |
| 4 | '000010000000000000000000000001000000000010' | 137472512002 | (5, 17, 30, 41) | (hp1-HP5) (hp3-HP3) (hp5-HP2) (ha1-HD1) | 3 | 0.8198 |
| 5 | '000010000000000010000100000000000000000010' | 137473032194 | (5, 17, 23, 41) | (hp1-HP5) (hp3-HP3) (hp4-HP2) (ha1-HD1) | 37 | 0.8198 |
| 6 | '000010000000000010001000000000000100100000' | 137473556768 | (5, 17, 22, 34, 37) | (hp1-HP5) (hp3-HP3) (hp4-HP1) (hp5-HD1) (ha1-HP2) | 3 | 0.5282 |
| 7 | '000010000000000100000000001010000000010000' | 137506103312 | (5, 16, 27, 29, 38) | (hp1-HP5) (hp3-HP2) (hp4-HD1) (hp5-HP1) (ha1-HP3) | 1 | 0.5282 |
| 8 | '100000000000000000000000000000000010000100' | 2199023255684 | (1, 35, 40) | (hp1-HP1) (hp5-HD2) (ha1-HP5) | 1 | 0.4274 |
| 9 | '100000001000000010000100000000000000000010' | 2207646875650 | (1, 9, 17, 25, 41) | (hp1-HP1) (hp2-HP2) (hp3-HP3) (hp4-HP4) (ha1-HD1) | 34 | 0.8702 |

**Supplementary Table S5** Detailed results of 100 effective experimental trials corresponding to Fig. 4(d), conducted under a constant $\sigma = 0.5$

| No. | Exp. Output in Binary format (42-bit) | Decimal | Output Graphs | Corresponding Docked PP pairs | Proportion (%) | Weights |
|---|---|---|---|---|---|---|
| 1 | '000000000000000000000000010001000000000001' | 69633 | (26, 30, 42) | (hp4-HP5) (hp5-HP2) (ha1-HD2) | 4 | 0.7694 |
| 2 | '000000000000000000000000010010000000000001' | 73729 | (26, 29, 42) | (hp4-HP5) (hp5-HP1) (ha1-HD2) | 8 | 0.7694 |
| 3 | '000000000000000000000100000000000011000000' | 524480 | (23, 35, 36) | (hp4-HP2) (hp5-HD2) (ha1-HP1) | 2 | 0.4274 |
| 4 | '000000000000000110000000000100010000000000000' | 402722816 | (14, 15, 26, 30) | (hp2-HD2) (hp3-HP1) (hp4-HP5) (hp5-HP2) | 1 | 0.2965 |
| 5 | '000000000000001000000000000000010000000001' | 1073750017 | (12, 29, 42) | (hp2-HP5) (hp5-HP1) (ha1-HD2) | 4 | 0.7694 |
| 6 | '000000000010000010000000000010000000000010' | 2181042178 | (11, 17, 30, 41) | (hp2-HP4) (hp3-HP3) (hp5-HP2) (ha1-HD1) | 2 | 0.8198 |
| 7 | '000000010000000000000000000000101000000' | 17179869344 | (8, 35, 37) | (hp2-HP1) (hp5-HD2) (ha1-HP2) | 1 | 0.4274 |
| 8 | '000010000000000010000100000000000000010' | 137473032194 | (5, 17, 23, 41) | (hp1-HP5) (hp3-HP3) (hp4-HP2) (ha1-HD1) | 6 | 0.8198 |
| 9 | '010000010000000000000000000001000000000001' | 1116691497473 | (2, 8, 33,42) | (hp1-HP2) (hp2-HP1) (hp5-HP5) (ha1-HD2) | 1 | 0.8198 |
| 10 | '100000000000000000000000000000010000100' | 2199023255684 | (1, 35, 40) | (hp1-HP1) (hp5-HD2) (ha1-HP5) | 2 | 0.4274 |
| 11 | '100000000000000000000010000000000000001' | 2199023321089 | (1, 26, 42) | (hp1-HP1) (hp4-HP5) (ha1-HD2) | 15 | 0.7694 |
| 12 | '100000000000000000001000000000000000100' | 2199025352708 | (1, 21, 40) | (hp1-HP1) (hp3-HD2) (ha1-HP5) | 12 | 0.4274 |
| 13 | '100000001000000000000000000000100001000' | 2207613190408 | (1, 9, 34, 39) | (hp1-HP1) (hp2-HP2) (hp5-HD1) (ha1-HP4) | 3 | 0.4778 |
| 14 | '100000001000000000000000000000100010000' | 2207613190416 | (1, 9, 34, 38) | (hp1-HP1) (hp2-HP2) (hp5-HD1) (ha1-HP3) | 1 | 0.4778 |
| 15 | '100000001000000000000000100000000001000' | 2207613222920 | (1, 9, 27, 39) | (hp1-HP1) (hp2-HP2) (hp4-HD1) (ha1-HP4) | 2 | 0.4778 |
| 16 | '100000001000000000010000000000000010000' | 2207617384464 | (1, 9, 20, 38) | (hp1-HP1) (hp2-HP2) (hp3-HD1) (ha1-HP3) | 1 | 0.4778 |
| 17 | '100000001000000010000000100000000000000010' | 2207646875650 | (1, 9, 17, 25, 41) | (hp1-HP1) (hp2-HP2) (hp3-HP3) (hp4-HP4) (ha1-HD1) | 35 | 0.8702 |

**Supplementary Table S6** Detailed results of 100 effective experimental trials corresponding to Fig. 4(e), conducted under a constant $\sigma = 0.2$

| No. | Exp. Output in Binary format (42-bit) | Decimal | Output Graphs | Corresponding Docked PP pairs | Proportion (%) | Weights |
|---|---|---|---|---|---|---|
| 1 | '000000000000000000000000000001010000000100' | 20484 | (28, 30, 40) | (hp4-HD2) (hp5-HP2) (ha1-HP5) | 1 | 0.4274 |
| 2 | '000000001000000000000000000000000011000000' | 8589934784 | (9, 35, 36) | (hp2-HP2) (hp5-HD2) (ha1-HP1) | 1 | 0.4274 |
| 3 | '000000010000000000000000000000101000000' | 17179869344 | (8, 35, 37) | (hp2-HP1) (hp5-HD2) (ha1-HP2) | 1 | 0.4274 |
| 4 | '000010000000000100000100000000000100100000' | 137473556768 | (5, 17, 22, 34, 37) | (hp1-HP5) (hp3-HP3) (hp4-HP1) (hp5-HD1) (ha1-HP2) | 1 | 0.5282 |
| 5 | '100000000000000000000000000000010000100' | 2199023255684 | (1, 35, 40) | (hp1-HP1) (hp5-HD2) (ha1-HP5) | 4 | 0.4274 |
| 6 | '100000000000000000000010000000000000001' | 2199023321089 | (1, 26, 42) | (hp1-HP1) (hp4-HP5) (ha1-HD2) | 37 | 0.7694 |
| 7 | '100000000000000000001000000000000000100' | 2199025352708 | (1, 21, 40) | (hp1-HP1) (hp3-HD2) (ha1-HP5) | 13 | 0.4274 |
| 8 | '100000001000000000000000000000100001000' | 2207613190408 | (1, 9, 34, 39) | (hp1-HP1) (hp2-HP2) (hp5-HD1) (ha1-HP4) | 1 | 0.4778 |
| 9 | '100000001000000000000000000000100010000' | 2207613190416 | (1, 9, 34, 38) | (hp1-HP1) (hp2-HP2) (hp5-HD1) (ha1-HP3) | 1 | 0.4778 |
| 10 | '100000001000000000000000100000000001000' | 2207613222920 | (1, 9, 27, 39) | (hp1-HP1) (hp2-HP2) (hp4-HD1) (ha1-HP4) | 12 | 0.4778 |
| 11 | '100000001000000000000010000010000010000' | 2207613223440 | (1, 9, 27, 33, 38) | (hp1-HP1) (hp2-HP2) (hp4-HD1) (hp5-HP5) (ha1-HP3) | 1 | 0.5282 |
| 12 | '100000001000000000010000000000000010000' | 2207617384464 | (1, 9, 20, 38) | (hp1-HP1) (hp2-HP2) (hp3-HD1) (ha1-HP3) | 1 | 0.4778 |
| 13 | '100000001000000010000000100000000000000010' | 2207646875650 | (1, 9, 17, 25, 41) | (hp1-HP1) (hp2-HP2) (hp3-HP3) (hp4-HP4) (ha1-HD1) | 26 | 0.8702 |

**Supplementary Table S7** Detailed results of 100 effective experimental trials corresponding to Fig. 5(d), conducted under a DSA schedule with $\sigma$ gradually reduced from 0.5 to 0.

| No. | Exp. Output in Binary format (42-bit) | Decimal | Output Graphs | Corresponding Docked PP pairs | Proportion (%) | Weights |
|---|---|---|---|---|---|---|
| 1 | '000000000000000000000000000101000000000100' | 20484 | (28, 30, 40) | (hp4-HD2) (hp5-HP2) (ha1-HP5) | 1 | 0.4274 |
| 2 | '000000000000000000000000100100000000000001' | 73729 | (26, 29, 42) | (hp4-HP5) (hp5-HP1) (ha1-HD2) | 4 | 0.7694 |
| 3 | '000000000001000000000000000001000000000001' | 1073745921 | (12, 30, 42) | (hp2-HP5) (hp5-HP2) (ha1-HD2) | 1 | 0.7694 |
| 4 | '000010000000000100000100000000000000000010' | 137473032194 | (5, 17, 23, 41) | (hp1-HP5) (hp3-HP3) (hp4-HP2) (ha1-HD1) | 1 | 0.8198 |
| 5 | '000100000000000000000000000001000010000000' | 274877909056 | (4, 31, 36) | (hp1-HP4) (hp5-HP3) (ha1-HP1) | 1 | 0.3325 |
| 6 | '010000010000000000000000000001000000000001' | 1116691497473 | (2, 8, 33, 42) | (hp1-HP2) (hp2-HP1) (hp5-HP5) (ha1-HD2) | 3 | 0.8198 |
| 7 | '100000000000000000000000000000000010000100' | 2199023255684 | (1, 35, 40) | (hp1-HP1) (hp5-HD2) (ha1-HP5) | 1 | 0.4274 |
| 8 | '100000000000000000000000000001000000000001' | 2199023256065 | (1, 33, 42) | (hp1-HP1) (hp5-HP5) (ha1-HD2) | 1 | 0.7694 |
| 9 | '100000000000000000000100000000000000000001' | 2199023321089 | (1, 26, 42) | (hp1-HP1) (hp4-HP5) (ha1-HD2) | 1 | 0.7694 |
| 10 | '100000000000000000001000000000000000000100' | 2199025352708 | (1, 21, 40) | (hp1-HP1) (hp3-HD2) (ha1-HP5) | 8 | 0.4274 |
| 11 | '100000001000000000000000100000000000001000' | 2207613222920 | (1, 9, 27, 39) | (hp1-HP1) (hp2-HP2) (hp4-HD1) (ha1-HP4) | 4 | 0.4778 |
| 12 | '100000001000000000010000000000000000010000' | 2207617384464 | (1, 9, 20, 38) | (hp1-HP1) (hp2-HP2) (hp3-HD1) (ha1HP3) | 1 | 0.4778 |
| 13 | '100000001000000010000000000001000000000' | 2207646744832 | (1, 9, 17, 34) | (hp1-HP1) (hp2-HP2) (hp3-HP3) (hp5-HD1) | 1 | 0.2965 |
| 14 | '100000001000000010000010000000000000000010' | 2207646875650 | (1, 9, 17, 25, 41) | (hp1-HP1) (hp2-HP2) (hp3-HP3) (hp4-HP4) (ha1-HD1) | 72 | 0.8702 |

**Supplementary Table S8** (a) Raw output graphs generated from 100 experimental trials of the p-computer at $\sigma = 0.5$ without post-processing, corresponding to Supplementary Table S5. (b) Graph configurations after applying the post-processing algorithm (greedy shrinking and 100 times local search expansion) to the raw outputs.

(a)

| No. | Output Graphs form P-computer (Seeds) | Proportion (%) | Weights |
|---|---|---|---|
| 1 | (26, 30, 42) | 4 | 0.7694 |
| 2 | (26, 29, 42) | 8 | 0.7694 |
| 3 | (23, 35, 36) | 2 | 0.4274 |
| 4 | (14, 15, 26, 30) | 1 | 0.2965 |
| 5 | (12, 29, 42) | 4 | 0.7694 |
| 6 | (11, 17, 30, 41) | 2 | 0.8198 |
| 7 | (8, 35, 37) | 1 | 0.4274 |
| 8 | (5, 17, 23, 41) | 6 | 0.8198 |
| 9 | (2, 8, 33, 42) | 1 | 0.8198 |
| 10 | (1, 35, 40) | 2 | 0.4274 |
| 11 | (1, 26, 42) | 15 | 0.7694 |
| 12 | (1, 21, 40) | 12 | 0.4274 |
| 13 | (1, 9, 34, 39) | 3 | 0.4778 |
| 14 | (1, 9, 34, 38) | 1 | 0.4778 |
| 15 | (1, 9, 27, 39) | 2 | 0.4778 |
| 16 | (1, 9, 20, 38) | 1 | 0.4778 |
| **17** | **(1, 9, 17, 25, 41)** | **35** | **0.8702** |

(b)

| No. | Graphs after Postprocessing | Proportion (%) | Weights |
|---|---|---|---|
| 1 | (26, 29, 42) | 8 | 0.7694 |
| 2 | (23, 35, 36) | 2 | 0.4274 |
| 3 | (14, 15, 26, 30) | 1 | 0.2965 |
| 4 | (11, 17, 30, 41) | 6 | 0.8198 |
| 5 | (7, 12, 16, 29) | 4 | 0.2965 |
| 6 | (5, 17, 23, 41) | 2 | 0.8198 |
| 7 | (5, 14, 15, 30) | 4 | 0.2965 |
| 8 | (2, 8, 33, 42) | 1 | 0.8198 |
| 9 | (2, 8, 32, 42) | 1 | 0.8198 |
| 10 | (1, 28, 31, 40) | 14 | 0.4778 |
| 11 | (1, 26, 42) | 15 | 0.7694 |
| 12 | (1, 9, 34, 39) | 3 | 0.4778 |
| 13 | (1, 9, 34, 38) | 1 | 0.4778 |
| 14 | (1, 9, 27, 33, 38) | 2 | 0.5282 |
| 15 | (1, 9, 20, 32, 38) | 1 | 0.5282 |
| **16** | **(1, 9, 17, 25, 41)** | **35** | **0.8702** |

**Supplementary Table S9** (a) Raw output graphs generated from 100 experimental trials of the p-computer under a DSA schedule with $\sigma$ gradually reduced from 0.5 to 0 without post-processing, corresponding to Supplementary Table S7. (b) Graph configurations after applying the post-processing algorithm (greedy shrinking and 100 times local search expansion) to the raw outputs.

**(a)**

| No. | Output Graphs | Proportion (%) | Weights |
|---|---|---|---|
| 1 | (28, 30, 40) | 1 | 0.4274 |
| 2 | (26, 29, 42) | 4 | 0.7694 |
| 3 | (12, 30, 42) | 1 | 0.7694 |
| 4 | (5, 17, 23, 41) | 1 | 0.8198 |
| 5 | (4, 31, 36) | 1 | 0.3325 |
| 6 | (2, 8, 33, 42) | 3 | 0.8198 |
| 7 | (1, 35, 40) | 1 | 0.4274 |
| 8 | (1, 33, 42) | 1 | 0.7694 |
| 9 | (1, 26, 42) | 1 | 0.7694 |
| 10 | (1, 21, 40) | 8 | 0.4274 |
| 11 | (1, 9, 27, 39) | 4 | 0.4778 |
| 12 | (1, 9, 20, 38) | 1 | 0.4778 |
| 13 | (1, 9, 17, 34) | 1 | 0.2965 |
| 14 | **(1, 9, 17, 25, 41)** | **72** | **0.8702** |

**(b)**

| No. | Output Graphs | Proportion (%) | Weights |
|---|---|---|---|
| 1 | (28, 30, 40) | 1 | 0.4274 |
| 2 | (26, 29, 42) | 4 | 0.7694 |
| 3 | (12, 30, 42) | 1 | 0.7694 |
| 4 | (11, 17, 30, 41) | 1 | 0.8198 |
| 5 | (5, 16, 31, 36) | 1 | 0.3829 |
| 6 | (2, 8, 33, 42) | 3 | 0.8198 |
| 7 | (1, 28, 31, 40) | 9 | 0.4778 |
| 8 | (1, 26, 42) | 1 | 0.7694 |
| 9 | (1, 9, 27, 33, 38) | 5 | 0.5282 |
| 10 | (1, 9, 20, 32, 38) | 1 | 0.5282 |
| 11 | **(1, 9, 17, 25, 41)** | **73** | **0.8702** |